\documentclass[footinbib,a4paper,aps,prb,reprint,twocolumn,preprintnumbers,amsmath,amssymb,10pt, superscriptaddress]{revtex4-2}
\usepackage{verbatim}
\usepackage{color}
\usepackage{tikz}
\usepackage{xr}
\usepackage[colorlinks=true,citecolor=blue,linkcolor=blue,urlcolor=blue]{hyperref}
\usepackage{braket}
\usepackage[normalem]{ulem}
\usepackage{upgreek}
\usepackage[percent]{overpic}
\usepackage{graphicx,xcolor}
\usepackage{dcolumn}
\usepackage{bm}
\usepackage[shortlabels]{enumitem}
\usepackage{bbold}

\newcommand{\Tr}{\text{Tr}}
\renewcommand{\rm}[1]{\textrm{#1}}

\definecolor{forestgreen}{rgb}{0.1, 0.6, 0.2}

\makeatletter
\renewcommand{\fnum@figure}{FIG. \thefigure}
\makeatother

\begin{document}

\title{Entanglement enhancement induced by noise in inhomogeneously monitored systems}
\author{Cristiano Muzzi}
\affiliation{SISSA — International School of Advanced Studies, via Bonomea 265, 34136 Trieste, Italy}
\affiliation{INFN, Sezione di Trieste, via Valerio 2, 34127 Trieste, Italy}
\author{Mikheil Tsitsishvili}
\affiliation{The Abdus Salam International Centre for Theoretical Physics (ICTP), Strada Costiera 11, 34151 Trieste, Italy}
\affiliation{SISSA — International School of Advanced Studies, via Bonomea 265, 34136 Trieste, Italy}
\affiliation{Institut f\"{u}r Theoretische Physik, Heinrich-Heine-Universit\"{a}t, D-40225 D\"{u}sseldorf, Germany}
\author{Giuliano Chiriac\`{o}}
\email{giuliano.chiriaco@dfa.unict.it}
\affiliation{Dipartimento di Fisica e Astronomia ``Ettore Majorana'', Universit\`{a}
di Catania, 95123 Catania, Italy}
\affiliation{INFN, Sez. Catania, I-95123 Catania, Italy}

\date{\today}

\begin{abstract}
We study how stronger noise can enhance the entanglement in inhomogeneously monitored quantum systems. We consider a free fermions model composed of two coupled chains -- a system chain and an ancilla chain, each subject to its own different noise -- and explore the dynamics of entanglement within the system chain under different noise intensities. Our results demonstrate that, contrary to the detrimental effects typically associated with noise, certain regimes of noise on the ancilla can significantly enhance entanglement within the system. Numerical simulations demonstrate the robustness of such entanglement enhancement across various system sizes and noise parameters. This enhancement is found to be highly dependent on the hopping strength in the ancilla, suggesting that the interplay between unitary dynamics and noise can be tuned to optimize the entanglement of a quantum system.
\end{abstract}

\maketitle

\section{Introduction}\label{Sec:Intro}

Quantum correlations and entanglement are fundamental properties of quantum systems, and play a crucial role in various quantum information and quantum computing tasks \cite{Horodecki2009,Plenio1998,amicoEntanglementManybodySystems2008a,preskillQuantumComputingNISQ2018}.

However, the physical realization of entangled states for practical purposes is often hampered by decoherence, which arises from the interaction of a quantum system with the surrounding environment \cite{Hornberger2009,  Weiss2011,Carmichael09,rivasEntanglementNonMarkovianityQuantum2010}. Decoherence and noise destroy entanglement, and are a significant obstacle to maintaining the quantum coherence required for effective quantum information processing \cite{DiVincenzo1999,Monroe2002,Knill2000,Lidar1998:DFS,Kwiat2000:DFSexp,Lidar2003:bookDFS,paladinoDecoherenceDueDiscrete2003}. Understanding and mitigating the effects of decoherence is thus crucial for the advancement of quantum technologies \cite{Molmer1999,Kim2010quantum,monroe2021programmable,schneider2012experimental,Ebadi20}. 
In the context of quantum systems, various strategies have been proposed to preserve quantum correlations in the presence of noise.

One class of strategies relies on quantum error correction (QEC) protocols \cite{Lidar1999:DFS,Plenio1997:DFS,Egan2020,Choi20,acharyaSuppressingQuantumErrors2023}, which encode the quantum information of a logical qubit into a larger Hilbert space using the redundancy of many physical qubits, and allow the detection and correction of errors. However, while QEC protocols protect quantum information from various kinds of errors, their implementation is technologically demanding. They typically require high-fidelity quantum gates and a large number of physical qubits to encode a single logical qubit -- leading to large overheads in qubit resources and complexity.

An alternative strategy is that of quantum error mitigation (QEM) which focuses on reducing the impact of errors in the quantum computation results using post-processing techniques \cite{caiQuantumErrorMitigation2023,temmeErrorMitigationShortDepth2017,endoPracticalQuantumError2018,kandalaErrorMitigationExtends2019,zhangErrormitigatedQuantumGates2020,bravyiMitigatingMeasurementErrors2021,loweUnifiedApproachDatadriven2021}. The main advantage of QEM is its applicability to near-term quantum devices, as it requires fewer qubits and less overhead compared to QEC, making it particularly attractive for current noisy intermediate-scale quantum devices. However, QEM techniques typically require precise knowledge of the noise characteristics, limiting their accuracy and their scalability to larger quantum systems.

In this work, we report a particularly intriguing phenomenon, which could lead to an alternative approach to protecting quantum coherences. We show that noise can induce an enhancement of entanglement in certain inhomogeneously monitored quantum systems. This counterintuitive phenomenon can be attributed to the monogamy of entanglement \cite{Dur:2000monogamy,Yang2006:monogamy,Giorgi:2011monogamy,Coffman2000,Ou2007} and to the interplay between memory, noise and the dynamics of the system \cite{fazio2024many,Nourmandipour2016,Hartmann2007,verstraete2009quantum,Diehl08,Sieberer16,Jin16,Maghrebi2016,FossFeig2017,Fink17,Fitzpatrick17,Flaschner18,Syassen08,Marino16,seetharam2021correlation}. These combined effects lead to an increase in entanglement, opposite to the decoherence usually caused by noise.

Indeed, the competition between a coherent drive and dissipative processes in many-body quantum systems has been extensively studied in recent years. It has been showed to lead to a plethora of non-equilibrium phases and to transitions between such phases. Important examples are dissipative phase transitions \cite{Sieberer16,Lee13,Jin16,Maghrebi2016,Overbeck2017,FossFeig2017,Jin18,Fink17,Fitzpatrick17,Flaschner18,Syassen08,Tomita17,Diehl08,Poletti12,Poletti13,Diehl08,Sieberer13,Marino16,Schiro16,Young20,paz2021drivendissipative,Marcuzzi14,seetharam2021correlation,Sierant2022dissipativefloquet}, measurement induced phase transitions \cite{Dhar16,Nahum2017, Li18,ChanUnitaryEntanglementDynamics2019,SzyniszewskiEntanglementTransitionWeakMeas:2019,Szyniszewski20, Li2019,Li20, Zhou19, Skinner2019, Bao2020, Jian20,Gullans2020, Gullans2020a,Gopalakrishnan20,Turkeshi2021a,Ippoliti21,Buchhold2021,Minato21,Block2021,Jian21,Cao19,Lang15,chiriacoDiagrammaticMethodManybody2023a,BoormanDiagnosticEntanglementDisorder2022,Romito_PhysRevA.110.022214_2024,nehra2024controllingmeasurementinducedphase,Gribben2024,kelly2024generalizingmeasurementinducedphasetransitions,leung2024entanglementoperatorcorrelationsignatures,paviglianiti2024breakdownmeasurementinducedphasetransitions,Shuo2024:PRBhybrid,Shuo2024:PRLhybrid}, transient dynamical phases \cite{Averitt02,Basov11,Morrison14,Fausti11,Mitrano16,Zong19,Kogar20,Disa20,Chiriaco18,Sentef17,Sun20,Chiriaco_PhysRevB.101.041105_2020}, non-equilibrium phases \cite{nakamuraElectricfieldinducedMetalMaintained2013,mazzaFieldDrivenMottGap2016,matthiesControlCompetingSuperconductivity2018,Chiriaco_PhysRevB.98.205152_2018_CDW,ZimmersPhysRevLett.110.056601,mcleodNanoimagingStraintunedStripe2021a,Nova19,Chiriaco_PhysRevB.102.085116_2020_CaRuO} and much more.

Here we present results that suggest that noise can also be exploited to enhance the effect of coherent processes. We investigate this phenomenon within a simple toy model of free fermions  \cite{Coppola2022,Mirlin2023,Alberton21,leung2023theoryfreefermionsdynamics,Kells2023:FF_Romito,favaNonlinearSigmaModels2023,SzyniszewskiDisorderedMonitoredFermions,poboikoMeasurementinducedTransitionsInteracting2024,starchl2024generalizedzenoeffectentanglement} consisting of two coupled chains: a system chain and an ancilla chain, each subjected to its own independent noise. We model the effect of noise (or any error or decoherence process) as arising from random projective measurements acting on the system via the Kraus representation. The model is similar to the one studied in a previous work \cite{tsitsishviliMeasurementInducedTransitions2024} coauthored by some of us. In that work only the ancilla chain was subject to random projective measurements of the occupation number. For certain values of the ancilla hopping strength, the entanglement within the system was enhanced with stronger ancilla noise. 

In this work we expand on such findings, by also considering noise acting on the system and characterizing the entanglement phases and the corresponding measurement induced transitions. Through extensive numerical simulations of quantum trajectories \cite{MolmerDalibard1993, DalibardMolmer1992,Plenio1998_quantum_jumps,Daley2014}, we demonstrate that stronger noise on the ancilla chain can significantly enhance the entanglement in the system chain. This enhancement consistently appears for various system sizes and noise parameters, but crucially depends on the hopping strength in the ancilla chain.

In particular, we find two regimes where the system exhibits different behaviors depending on the ancilla dynamics. In the slow dynamics (small hopping) regime, any noise on the ancilla is detrimental to entanglement in the system. In the fast dynamics (large ancilla hopping) regime, the entanglement in the system is enhanced by noise acting on the ancilla, which protects the entangled phase, even in the presence of strong noise acting on the system. Such effect is a consequence of both the monogamy of entanglement and memory effects introduced by the fast dynamics of the ancilla, which leads the system to experience an effective (correlated) non-Markovian noise \cite{breuerColloquiumNonMarkovianDynamics2016,paladino$mathbsf1Mathbsfitf$Noise2014}. This is related to recent findings that non-Markovian noise \cite{chiriacoDiagrammaticMethodManybody2023a,SzyniszewskiDisorderedMonitoredFermions,tsitsishviliMeasurementInducedTransitions2024,niroulaErrorMitigationThresholds2024}, or long-range correlated noise \cite{Russomanno_PhysRevB.108.104313_2023,Piccitto_SciPostPhysCore.6.4.078_2023} can enhance the entanglement; indeed, entanglement creation via correlated dissipation was proposed and realized \cite{CiracPolzik:2011,Polzik_PhysRevLett.107.080503_2011} in few-body systems.

Our results suggest that the interplay between unitary dynamics and noise can be finely tuned to enhance entanglement, offering new insights into this complex topic. These findings may open new avenues for exploiting noise to improve the reliability and efficiency of quantum information processes.

The rest of the paper is organized as follows. In Section \ref{Sec:Model} we describe the model of the system, the evolution protocol,  how to calculate the entanglement negativity and the methods employed in the numerical simulations. In Section \ref{Sec:Results} we present the results of our work. Section \ref{Sec:Susce} focus on a more qualitative analysis of the negativity and of the susceptibility of entanglement with respect to the ancilla noise. Section \ref{Sec:scaling} presents the results of a scaling analysis on the behavior of the negativity. Section \ref{Sec:NoiseTrans} and \ref{Sec:HoppingTrans} show how the various entanglement phases emerge depending on the value of the ancilla noise strength and of the ancilla hopping. Finally Section \ref{Sec:Conclusions} presents our conclusions and outlook.

\section{Model}\label{Sec:Model}

\begin{figure}[t!]
    \centering
    \includegraphics[width=0.96\columnwidth]{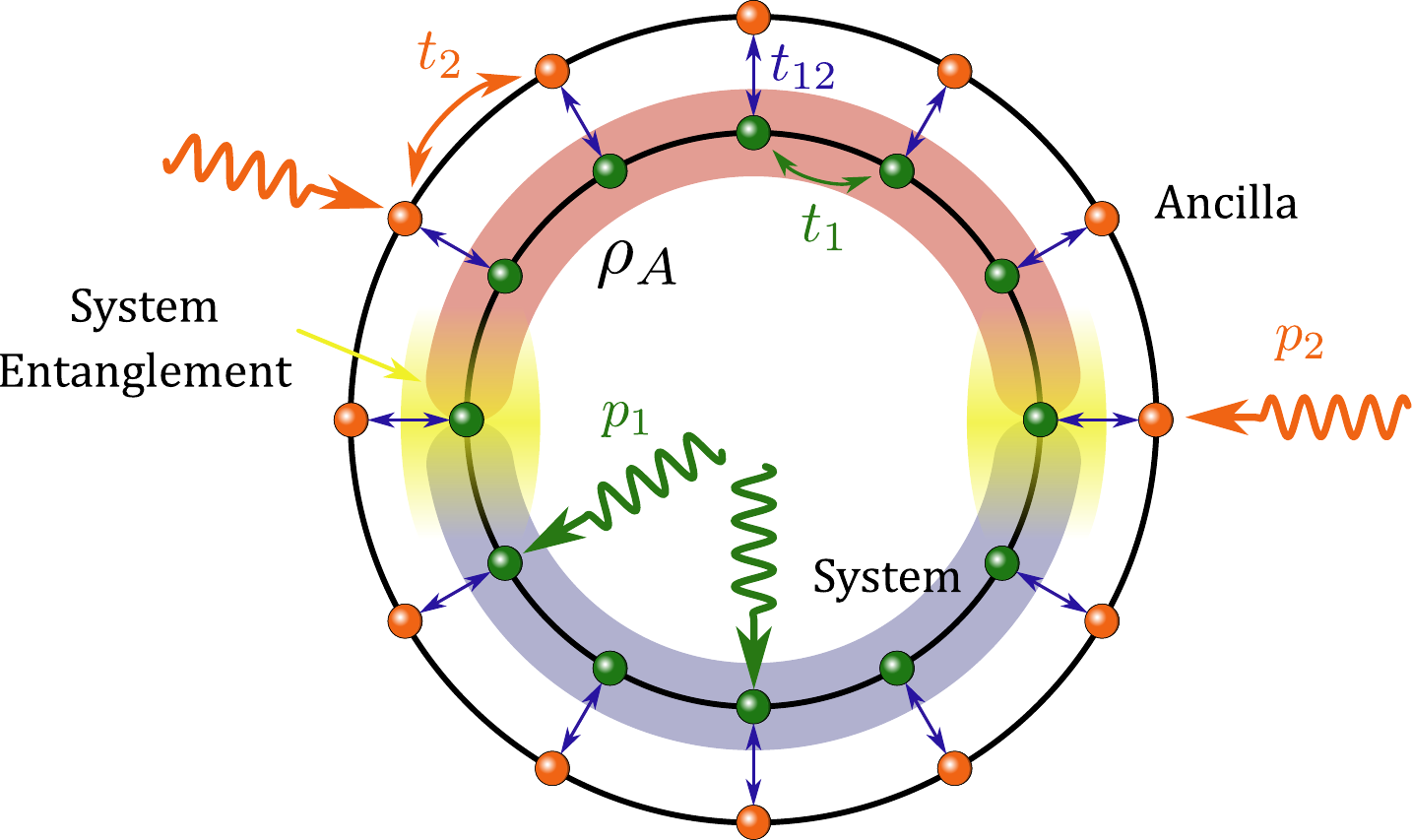}
    \caption{Sketch of the investigated fermionic model. The green (orange) spheres indicate the inner (outer) chain of fermions, representing respectively the system and the ancilla. Fermions can hop with strength $t_1$ within the system, $t_2$ within the ancilla and $t_{12}$ between system and ancilla. The wavy lines reprsent noise on the system and ancilla. In order to find the system entanglement, the system is partitioned into two parts $A$ and $B$, and the entanglement between the two is studied.}
    \label{fig:ladder}
\end{figure}

We consider two coupled free fermion chains, each one composed of $L$ sites, as depicted in Fig. \ref{fig:ladder}. The system is described by the quadratic Hamiltonian
\begin{equation}\label{Eq:modelH}
        \hat{H} = \sum_{i,\sigma}t_{\sigma}\hat{c}^{\dagger}_{i,\sigma}\hat{c}^{\phantom{\dagger}}_{i+1,\sigma} +
        t_{12}\sum^{L}_{i=1}\hat{c}^{\dagger}_{i,1}\hat{c}^{\phantom{\dagger}}_{i,2} + \text{h.c.},
\end{equation}
where $\hat{c}^{\dagger}_{j,\sigma},\hat{c}^{\phantom{\dag}}_{j,\sigma}$, denote, respectively, the fermionic creation and annihilation operators on site $i$ of chain $\sigma=1,2$. On each chain, periodic boundary conditions ($\hat{c}_{L+n,\sigma} = \hat{c}_{n,\sigma}$) are imposed. The parameters $t_1$ and $t_2$ denote the hopping amplitude in chain $1$ and $2$, respectively; the fermions can hop between the chains with amplitude $t_{12}$. Due to this inter-chain coupling this model is often referred to as a free-fermionic \textit{ladder}, which is sketched in Fig.~\ref{fig:ladder}.

\begin{figure}[t!]
    \centering
    \includegraphics[width=\columnwidth]{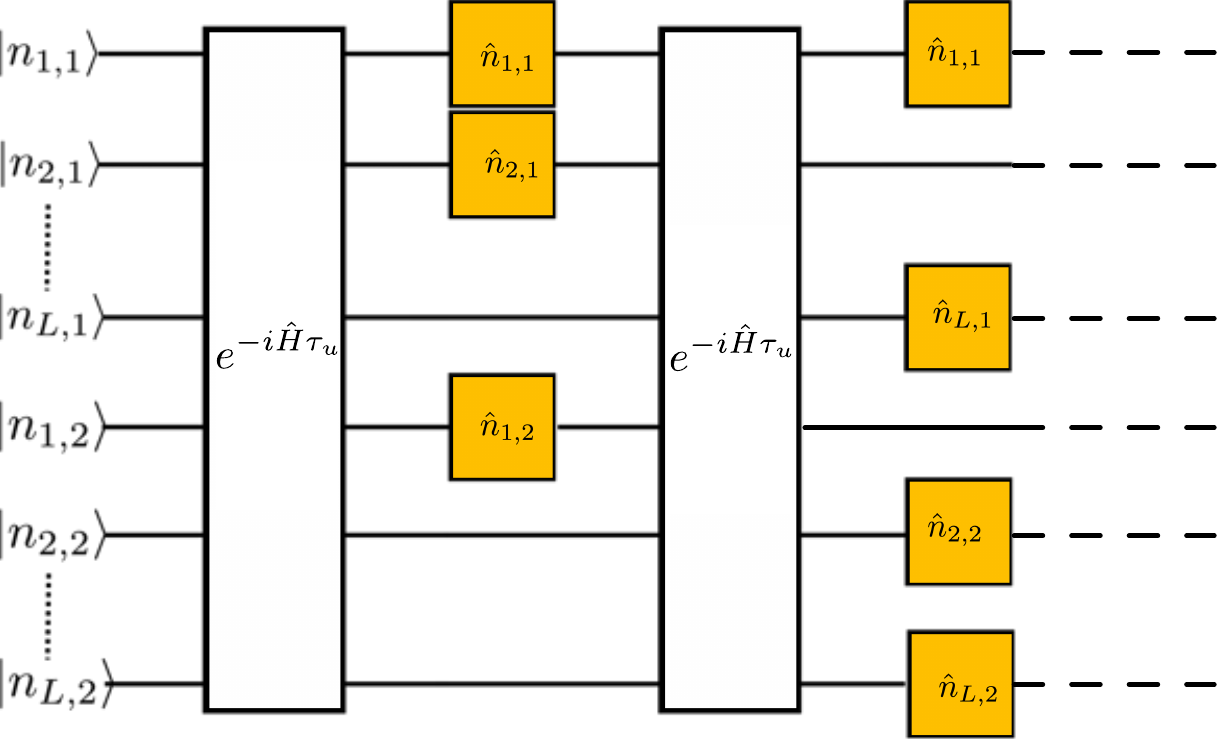}
    \caption{Sketch of the evolution of the ladder. The state is initialized in a factorized state at half filling, with random occupation number $n_{i,1/2}$. Such state is then evolved for a time $\tau_u$ with the quadratic Hamiltonian in Eq. \eqref{Eq:modelH}, and then subject to random measurements of the occupation number, modelling the action of the noise. This evolution cycle is repeated until the steady state is reached.}
    \label{fig:Circ_Fermions}
\end{figure}

In our study we model the effect of noise via the action of random projective measurements. We consider the hamiltonian evolution of the system described by Eq.(\ref{Eq:modelH}) interspersed with projective measurements of the particle number $n_{i,\sigma}=c^{\dagger}_{i,\sigma}c^{\phantom{\dag}}_{i,\sigma}$. The probability of performing the measurement on chain $\sigma$ is denoted by $p_\sigma$, and is a measure of the noise strength. Ultimately, we are interested in the entanglement properties of chain 1, which can be extracted by tracing out the degrees of freedom of chain $2$. For this reason we will refer to chain $1$ as the \textit{system} and chain $2$ as the \emph{ancilla}. The ancilla is measured with probability $p_2$, while the system is measured with probability $p_1$. The dynamics consists of alternating unitary evolution and projective measurements (see Fig. \ref{fig:Circ_Fermions}) and is formulated as follows \cite{Coppola2022,tsitsishviliMeasurementInducedTransitions2024}: 

\begin{enumerate}
    \item The system and the ancilla total state is initiated at half-filling in a random product state, i.e. $\ket{\Psi(0)}= \ket{n_{1,1},...,n_{L,1},n_{1,2},...,n_{L,2}}$, where $n_{i,\sigma}=0,1$ is the expectation value of the occupation number.
    \item The state $\ket{\Psi(0)}$ evolves to $\ket{\Psi(\tau_{u})}$ unitarily for a time $\tau_u$ according to the evolution generated by the Hamiltonian Eq. \eqref{Eq:modelH}.
    \item The state $\ket{\Psi(\tau_{u})}$ is subject to projective measurements of the local particle number $\hat{n}_{i,\sigma}$ on each site. It is assumed that the measurements are instantaneous and act on $\ket{\Psi(\tau_{u})}$ according to the Kraus representation, see Appendix \ref{App:Evolution}.
    \item The sequence of unitary evolution (step 2) and projective measurements (step 3) is performed $N_{st}$ times, until the desired temporal depth of evolution $\tau_{st}=N_{st}\tau_{u}$ is reached. At this point, the evolution of the ladder is terminated and the final state $\ket{\Psi(\tau_{st})}$ is extracted.
\end{enumerate}

The time evolved state of the ladder is easily obtained. Let us first note that the Hamiltonian Eq.(\ref{Eq:modelH}) can be diagonalized in Fourier space. Introducing the Fourier modes $\hat{c}_{k,\sigma}=\sum_j e^{-ijk}\hat{c}_{j,\sigma}/\sqrt{L}$ and the Nambu spinors $\hat{\psi}^{\dagger}_k\equiv (\hat{c}^{\dagger}_{k,1},\hat{c}^{\dagger}_{k,2})$, we can write the Hamiltonian as 
\begin{gather}
    \hat{H} = \sum_{k} \hat{\psi}^\dagger_k \hat{H}_k \hat{\psi}_k;\\
\label{H_k}
    \hat H_k = \begin{pmatrix}
        2t_1\cos k & t_{12}\\
        t_{12} & 2t_2 \cos k
    \end{pmatrix}
\end{gather}
The state of the ladder after a time $\tau_u$ is given by $\ket{\Psi(\tau_u)}=\hat{U}\ket{\Psi(0)}$, where the unitary evolution operator for a time $\tau_u$ factorizes as $\hat{U}=\otimes_k \hat{U}_k$, with $\hat{U}_k=e^{-i\tau_u\hat{H}_k}$, explicitly given by:
\begin{equation}
\label{U_k}
    \begin{split}
        &\hat{U}_{k} = e^{-it\cos(k\tau_u)}\biggr[\hat{\mathbb{I}}\cos\left(\sqrt{t^2_{12}+\delta^2\cos^2k}\tau_{u}\right) 
        \\
        &-i\frac{t_{12} \hat{\sigma}^{x}+\delta\cos k \hat{\sigma}^{z}}{\sqrt{t^2_{12}+\delta^2\cos^2k}}\sin\left(\sqrt{t^2_{12}+\delta^2\cos^2k}\tau_{u}\right)\biggr],
    \end{split}
\end{equation}
where $t=t_1+t_2$, $\delta=t_{1}-t_{2}$, and the identity $\hat{\mathbb{I}}$ and the Pauli matrices $\hat{\sigma}^{x,z}$ operate in the Nambu space.

We observe that at $\delta=0$ the evolution is periodic in $t_{12}$ for every $k$. In particular, if $t_{12}\tau_u=\pi+\pi n$ ($n \in \mathbb{Z}$), the evolution of the two chains is effectively decoupled, while for $t_{12}\tau_{u}=\pi/2+\pi n$ ($n \in \mathbb{Z}$) the coupling between the two chains is maximal. Choosing $t_{12}\tau_u=\pi/2$ would maximize memory effects from the ancilla chain on the system chain.

After each unitary evolution cycle for a time $\tau_u$, the local particle occupation numbers $n_{i,\sigma}$ are measured with probability $p_\sigma$ on each site. The state of the ladder after the measurement is given by an eigenstate of the particle occupation number operators. 

After repeating the unitary and measurement cycle for $N_{st}$ times, the ensemble of the two chains is described by the density matrix of a pure state $\rho = \ket{\Psi(\tau_{st})}\bra{\Psi(\tau_{st})}$. Tracing over the degrees of freedom of the ancilla yields the reduced density matrix of the system $\rho_1=\text{Tr}_2 \rho$. For $p_{2} < 1$ the system can generally be entangled with the ancilla and its state may be mixed, so that the entanglement entropy fails to be a valid witness of quantum entanglement within the system \cite{Plenio1998,Donald2002,Plenio2006introduction, Horodecki2009}, since it would also take into account classical correlations. 
The quantum entanglement between two partitions -- one partition A containing $l_{A}$ adjacent sites and its complement B with the remaining $l_{B}=L-l_{A}$ sites -- of a system in a mixed state can be faithfully quantified by the (logaritmic) fermionic negativity. The negativity is equivalently obtained by performing a partial time-reversal transformation \cite{Shapourian2017,Shapourian_2019} or a fermionic twisted partial transpose \cite{Ruggiero2019:NegativityFF,Murciano2022:Negativity,CalabreseNegativity2012} of the density matrix.  In our work we employ the first method

\begin{equation}
\label{eq:negativity_def}
    \mathcal{\mathcal{E}}=\log \text{Tr}|\rho^{R_A}_1| =\sqrt{\rho^{R_A}_1(\rho^{R_A}_1)^{\dagger}}.
\end{equation}
where $\rho^{R_A}_1$ denotes the partial time reversal, whose action is explicitly written in the Appendix \ref{App:Negativity}.

\begin{figure*}[t!]
    \centering
    \includegraphics[width=\textwidth]{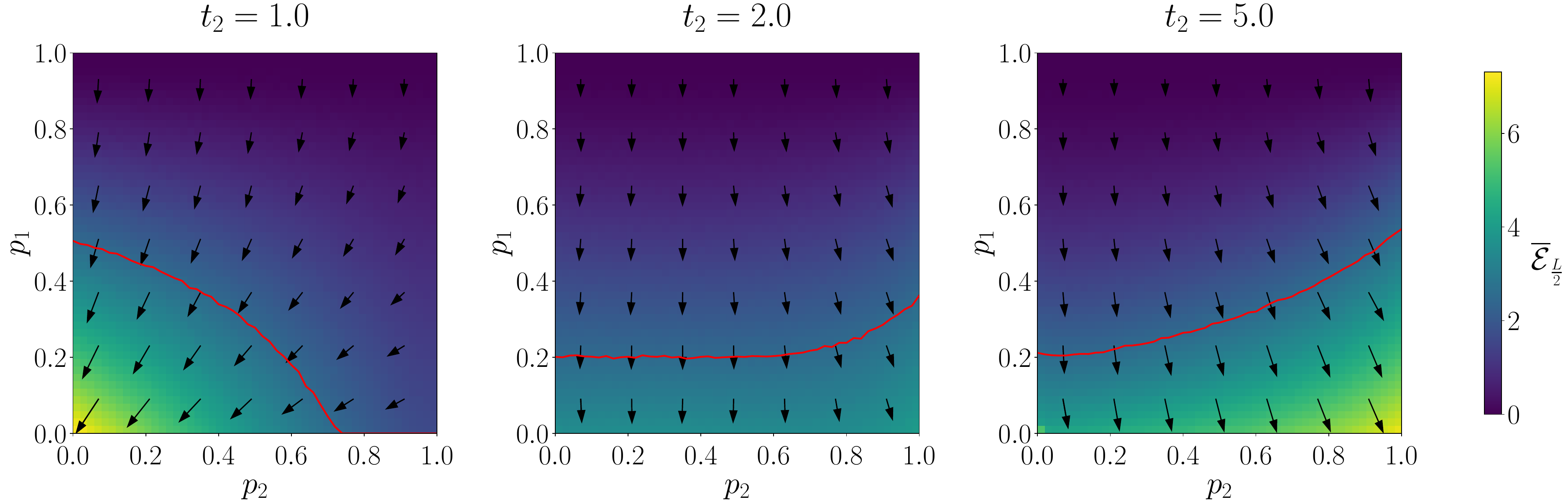}
\put(-497,155){(a)} 
\put(-339,155){(b)} 
\put(-183,155){(c)} 
    \caption{Colormap of the negativity in the inner chain for a half bipartition $\overline{\mathcal{E}}_{\frac L2}$, plotted as function of the measurement probability in the outer $p_2$ and inner chain $p_1$, for $L=32$ and for selected values of $t_2$. The red contour lines indicate where $\overline{\mathcal{E}}_{\frac L2}=2.5$, chosen to show qualitatively the low and high entanglement regions;} the arrows express the gradient of the negativity.
    \label{fig:NegL32}
\end{figure*}

In the present case both the unitary dynamics (generated by a quadratic hamiltonian) and dissipative dynamics (measurement of fermion number) preserve the Gaussianity of the state \cite{Coppola2022,SuraceSciPostPhysLectNotes.54,Cao19}, which allows to extract entanglement measures from the two-point correlation function $\mathcal{D}_{ij,\sigma \sigma'}(\tau)= \bra{\Psi(\tau)}\hat{c}^\dagger_{i,\sigma}\hat{c}_{j,\sigma'} \ket{\Psi(\tau)}$ at time $\tau$. In particular, this property allows us to extract the bipartite fermionic negativity directly from $\mathcal{D}_{ij,\sigma \sigma'}(\tau)$ \cite{Turkeshi22:NegativityFF,Shapourian_2019,Ruggiero2019:NegativityFF} (see App. \ref{App:Negativity}), providing a significant numerical advantage compared to the simulation of the full density matrix $\rho$. Indeed, it is generally known that, for gaussian states, quantities like the entanglement entropy can be computed simply knowing the two-point correlation matrix \cite{Chung_2001,Cheong,Peschel_2003,Peschel_2009}, as it completely characterizes the reduced density matrix of the system. The negativity is not so straightforward to compute, as the standard partial transpose \cite{Peres, VidalNegativity, Horodecki2009} does not preserve gaussianity for fermionic systems \cite{eisler2015partial,Shapourian2017,Shapourian2019:NegativityFF}. However, both partial time-reversal transformation Eq. \eqref{eq:negativity_def} and the twisted partial transpose \cite{Ruggiero2019:NegativityFF} preserve gaussianity, and thus the corresponding fermionic negativities can be extracted easily from correlation functions. Indeed both these measures of entanglement were used to characterize mixed state entanglement in driven and open fermionic quantum systems \cite{Turkeshi22:NegativityFF,Alba_2023,Murciano2022:Negativity}.

The dynamics of the monitored free fermionic ladder is investigated within the quantum trajectory approach \cite{MolmerDalibard1993, DalibardMolmer1992,Plenio1998_quantum_jumps,Daley2014}. We simulate the evolution of the ladder for a number $N_{traj}$ of different times for a random $\ket{\Psi(0)}$. The stocasticity of the measurement outcomes along each trajectory gives rise to an ensemble of different quantum trajectories. For large enough times $\tau>\tau_{st}$, the negativity that evolves along a single trajectory saturates to a steady state (independent of the initial conditions \cite{Turkeshi22:NegativityFF,tsitsishviliMeasurementInducedTransitions2024}) value around which it fluctuates. For a given trajectory $\alpha$ we define the steady state average of the negativity as
\begin{equation}
   \langle \langle  \mathcal{E}^\alpha \rangle \rangle  = \frac{1}{m}\sum^m_{s=1}\mathcal{E}^{(\alpha)}(\tau_{st}+s\tau_u),
\end{equation}
which is performed over the next $m$ steps after $\tau_{st}$ \footnote{We note that fluctuations around the steady state are uncorrelated for each trajectory, so that averaging over trajectories gets rid of these fluctuations, without needing to average over the last time steps for each trajectory.}. A safe estimate for $\tau_{st}$ for the system sizes considered in this work is $\tau_{st}=150$ for $L\leq64$ and $\tau_{st}=250$ for $L>64$ \cite{tsitsishviliMeasurementInducedTransitions2024}. For concreteness, in our study we have used $m=5$, $\tau_{st}=250$ and $N_{traj}=150$.

By averaging the stationary value of the negativity over the ensemble of trajectories, we obtain the trajectory averaged steady state value of the negativity 
\begin{equation}
    \overline{\mathcal{E}} = \frac{1}{N_{traj}}\sum^{N_{traj}}_{\alpha=1} \langle \langle  \mathcal{E}^\alpha \rangle \rangle
\end{equation}
Below, we focus on the properties of the trajectory averaged fermionic negativity $\overline{\mathcal{E}}_{A}$.

\section{Results}\label{Sec:Results}

We want to understand how the noise acting on the ancilla and on the system impacts the entanglement in the system, for various regimes of the dynamics of the ancilla $t_2$. Since the tunneling amplitude $t_{2}$ dictates how quickly a particle delocalizes in the ancilla, and thus how quickly entanglement spreads in the ancilla, we can use the $t_{2}/t_{1}$ ratio to characterize how fast the ancilla dynamics is relative to the system dynamics. For convenience, in the following we fix $\tau_u=1$ and $t_1=1$; the value of $t_1$ sets the unit of measure of all other tunneling amplitudes. Moreover, we set $t_{12}=\pi/2$ in order to achieve maximal coupling between the chains.

\begin{figure*}[t!]
    \centering
    \includegraphics[width=\textwidth]{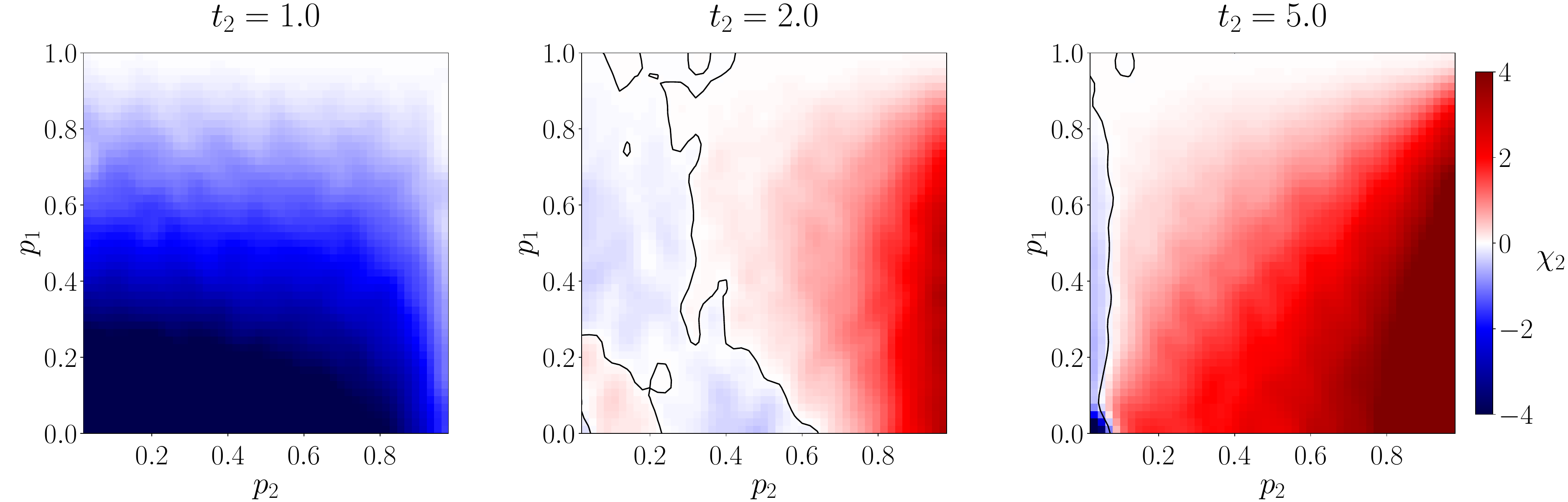}
\put(-495,155){(a)} 
\put(-332,155){(b)} 
\put(-169,155){(c)} 
    \caption{Colormap of the derivative of the negativity $\overline{\mathcal{E}}_{\frac L2}$ for $L=32$ with respect to $p_2$, for selected values of $t_2$. The negativity is plotted as function of the measurement probability in the outer chain $p_2$ and the measurement probability in the inner chain $p_1$. The white color in the color map corresponds to regions where the derivative vanishes and the susceptibility of entanglement changes sign.}
    \label{fig:Gradient}
\end{figure*}

\subsection{Negativity and entanglement susceptibility}\label{Sec:Susce}

We first study the negativity as function of $p_1$ and $p_2$ for a fixed chain size $L=32$ and for a bipartition with size $l_A=L/2$, for a chain of size $L=32$ and several values of $t_2$. We plot $\overline{\mathcal{E}}_A(L/2)$, see Fig.~\ref{fig:NegL32}; from the colormap, we can roughly gauge whether the two partitions in the system are highly entangled or not. 

For values of $t_2$ comparable with $t_1$, e.g. $t_2 =1.0$ (see panel a), the negativity is maximum for $p_1=p_2=0$, i.e. when both the system and the ancilla are not subject to any noise. In this regime, increasing either $p_1$ or $p_2$ lowers the value of the negativity, meaning that measurements on the system or on the ancilla lead to a decrease of the entanglement between partitions of the system. This is the expected behavior: stronger noise yields smaller entanglement.

However, as $t_2$ is increased, for example $t_2=2.0$ (see Fig~\ref{fig:NegL32} panel b), the negativity does not decrease for larger values of $p_2$, but actually slightly increases and reaches a maximum near $p_2=1$. This is a counter-intuitive trend, which is clearly evident for $t_2\gg t_1$, e.g. $t_2=5.0$, see Fig.~\ref{fig:NegL32} panel c. For such value of $t_2$, the negativity is largest for $p_1=0, p_2=1$, meaning that the entanglement between the two partitions of the system is maximal when the ancilla is always measured. Thus, as the dynamics in the ancilla gets faster, the entanglement in the system can be increased by applying a stronger noise on the ancilla.

Such remarkable effect is better characterized by introducing the entanglement susceptibility to measurement rate of the ancilla
\begin{equation}
    \chi_2(p_1,p_2,t_2)=\frac{d\overline{\mathcal{E}}_{\frac L2}}{dp_2},
\end{equation}
which we study as function of $p_1$ and $p_2$ for the same values of $t_2$ as in Fig.~\ref{fig:NegL32}, and for a system of size $L=32$. The behavior of the susceptibility is shown in Fig. \ref{fig:Gradient}. 

In agreement with our previous observation, we notice that for low values of $t_2$, $\chi_2$ is always non positive in the whole phase diagram. In this regime, increasing the measurement rate of the ancilla leads to a decrease of the system entanglement. As $t_2$ is increased, however, the suceptibility becomes positive and higher in magnitude for high values of $p_2$, again signaling an increase in entanglement within the system when the ancilla is subject to stronger noise, provided that the system is not measured more frequently than the ancilla ($p_1<p_2$).

This effect is highly dependent on the value of the hopping in the ancilla. An intuitive explanation is due to the principle of monogamy of entanglement, which states that only a finite amount of entanglement can be shared among different partitions of a system. In our case, we consider partitions of the ancilla and the $A$ and $B$ partitions of the system. When $t_2\gg t_1$, entanglement spreading within the ancilla is more efficient than within the system; by virtue of the monogamy of entanglement the negativity between $A$ and $B$ is suppressed because most of entanglement is shared between partitions of the ancilla; we also point out that also entanglement between the system and ancilla themselves is suppressed because $t_2\ll t_{12}$. However, measuring the ancilla with stronger $p_2$ decreases its correlations and makes more entanglement available to be shared between $A$ and $B$, explaining why the negativity increases with $p_2$. This effect does not appear for $t_2\approx t_1$, since in this regime the entanglement content within the ancilla and within the system is comparable and thus measuring the ancilla does not free up much entanglement, but rather destroys correlations within the system in an indirect way via the coupling $t_{12}$.

Another possible factor contributing to this effect is the non-Markovianity introduced by the ancilla dynamics. In fact, the ancilla acts as a filter on the $p_2$ noise, which is then experienced by the system as a noise with memory effects. In particular, the non-Markovianity of the system evolution has been quantified in Ref. \cite{tsitsishviliMeasurementInducedTransitions2024} and has been showed to be larger for higher values of $t_2$. Large memory effects lead to a backflow of information from the environment into the system, leading to a revival of correlations \cite{chiriacoDiagrammaticMethodManybody2023a} and can effectively act as antinoise \cite{niroulaErrorMitigationThresholds2024}. Moreover, correlated noise in space \cite{Piccitto_SciPostPhysCore.6.4.078_2023,Russomanno_PhysRevB.108.104313_2023,LumiaPhysRevResearch.6.023176_2024} has been shown to substantially increase entanglement, depending on the correlation length of the noise. A similar enhancement was found for noise arising from certain regimes of weak spatial and temporal disorder in \cite{BoormanDiagnosticEntanglementDisorder2022,SzyniszewskiDisorderedMonitoredFermions}. In our model, the noise is correlated in time due to the ancilla dynamics, with $t_2$ acting as a control parameter of the correlation time/length. The results found so far, then suggest that temporal correlations induced by non-Markovianity have a similar effect to spatial correlations of the noise, by increasing the entanglement content in the system.

\begin{figure*}[t!]
    \centering
    \includegraphics[width=\textwidth]{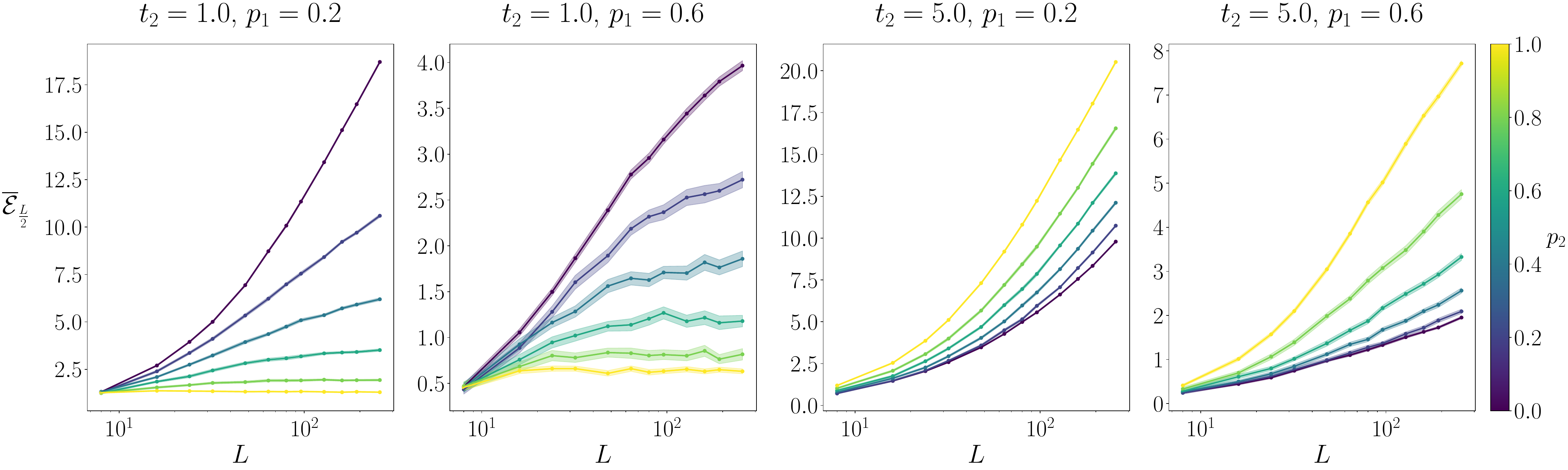}
\put(-495,145){(a)} 
\put(-375,145){(b)} 
\put(-252,145){(c)} 
\put(-138,145){(d)} 
    \caption{Plot of the negativity in the inner chain for half bipartition $\overline{\mathcal{E}}_{\frac L2}$ as function of $L$ (in logarithmic scale), for several values of the measurement probability in the outer chain $p_2\in\{0.0,0.2,0.4,0.6,0.8,1.0\}$ as indicated by the colorbar legend. Each plot represent a different value of $t_2\in\{1,5\}$ and of the measurement probability in the inner chain $p_1\in\{0.2,0.6\}$. The negativity at small $t_2$ and $p_1$ in (a) exhibits a different behavior as function of $p_2$: it scales logarithmically with $L$ for small $p_2$ and saturates to a constant for large $L$ as $p_2$ increases. On the other hand, at large $t_2$ (c)-(d) the negativity seems to always behave logarithmically no matter the value of $p_2$, actually displaying an enhancement with larger $p_2$.}
    \label{fig:Neg_vs_L}
\end{figure*}
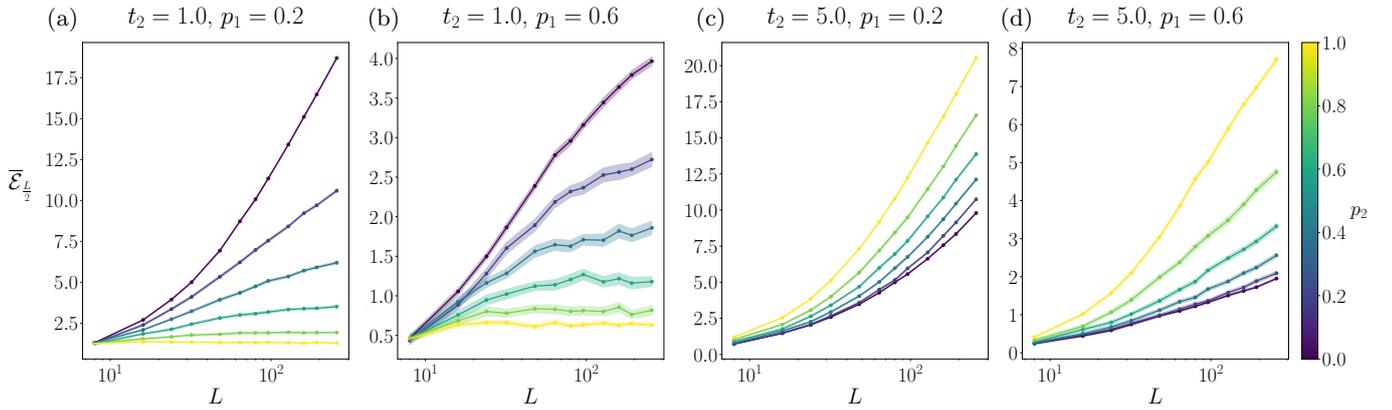

\subsection{Scaling analysis}\label{Sec:scaling}

The color maps in Fig. \ref{fig:NegL32} only show how much entanglement is contained in the system, but do not determine if the system entanglement is in an area law phase, a logarithmic law phase, or a volume law phase. In order to quantify this property we need to perform a study of the scaling properties of the negativity with the system size $L$. We calculate the negativity $\overline{\mathcal{E}}_{L/2}$ for several system sizes $L\in(8,16,24,32,48,64,80,96,128,160,192,256)$ as function of $p_2$. We investigate four different regimes: regime of weak ($p_1=0.2$) and strong $(p_1=0.6)$ noise on the system; regime of slow ($t_2=1.0$) and fast ($t_2=5.0$) dynamics of the ancilla. The results are reported in Fig. \ref{fig:Neg_vs_L} for $p_2\in(0,0.2,0.4,0.6,0.8,1)$.

When the dynamics of the ancilla is slow, and the measurement rate of the system is low ($t_2=1.0,p_1=0.2$, panel a) we observe two different behaviors. At small $p_2$ the entanglement is clearly logarithmic in $L$, and at larger values of $p_2$ it saturates to a constant for large enough sizes, indicating an area law. This is expected for weakly measured non-interacting systems \cite{Alberton21,BoormanDiagnosticEntanglementDisorder2022,tsitsishviliMeasurementInducedTransitions2024}; while analytic calculations \cite{Mirlin2023} suggest that the logarithmic scaling is a finite size effect, our model has a different underlying unitary dynamics, and it may still display a logarithmic phase. As the measurement rate of the system $p_1$ grows ($t_2=1.0, p_1=0.6$, panel b), the system appears instead to always be in an area law -- the results at $p_2=0$ suggest a saturation at large system sizes with finite size logarithmic corrections. Such behavior of the system at small values of $t_2$ is reasonable and expected: there is no entanglement monogamy effect, and an increase in either $p_1$ or $p_2$ decreases the entanglement, causing the system to switch from logarithmic to area law.

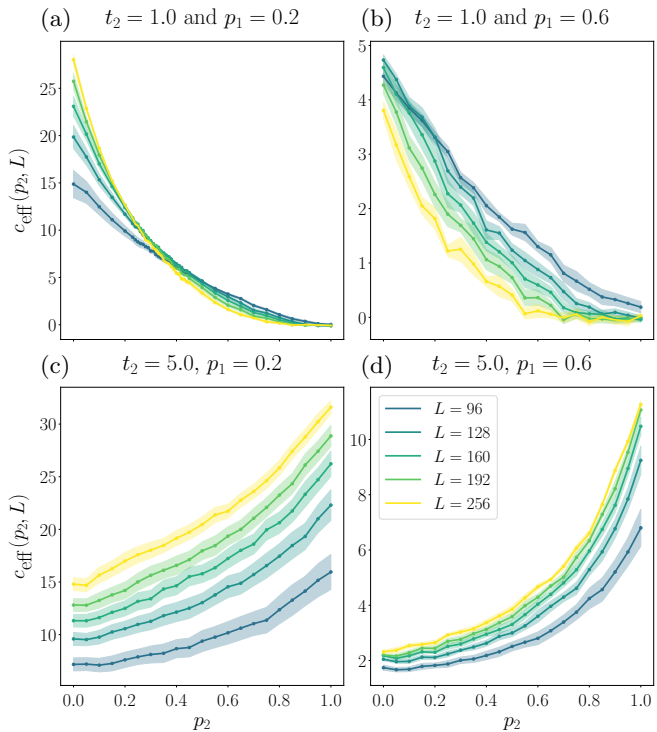
\begin{figure}[t!]
    \centering
    \includegraphics[width=\columnwidth]{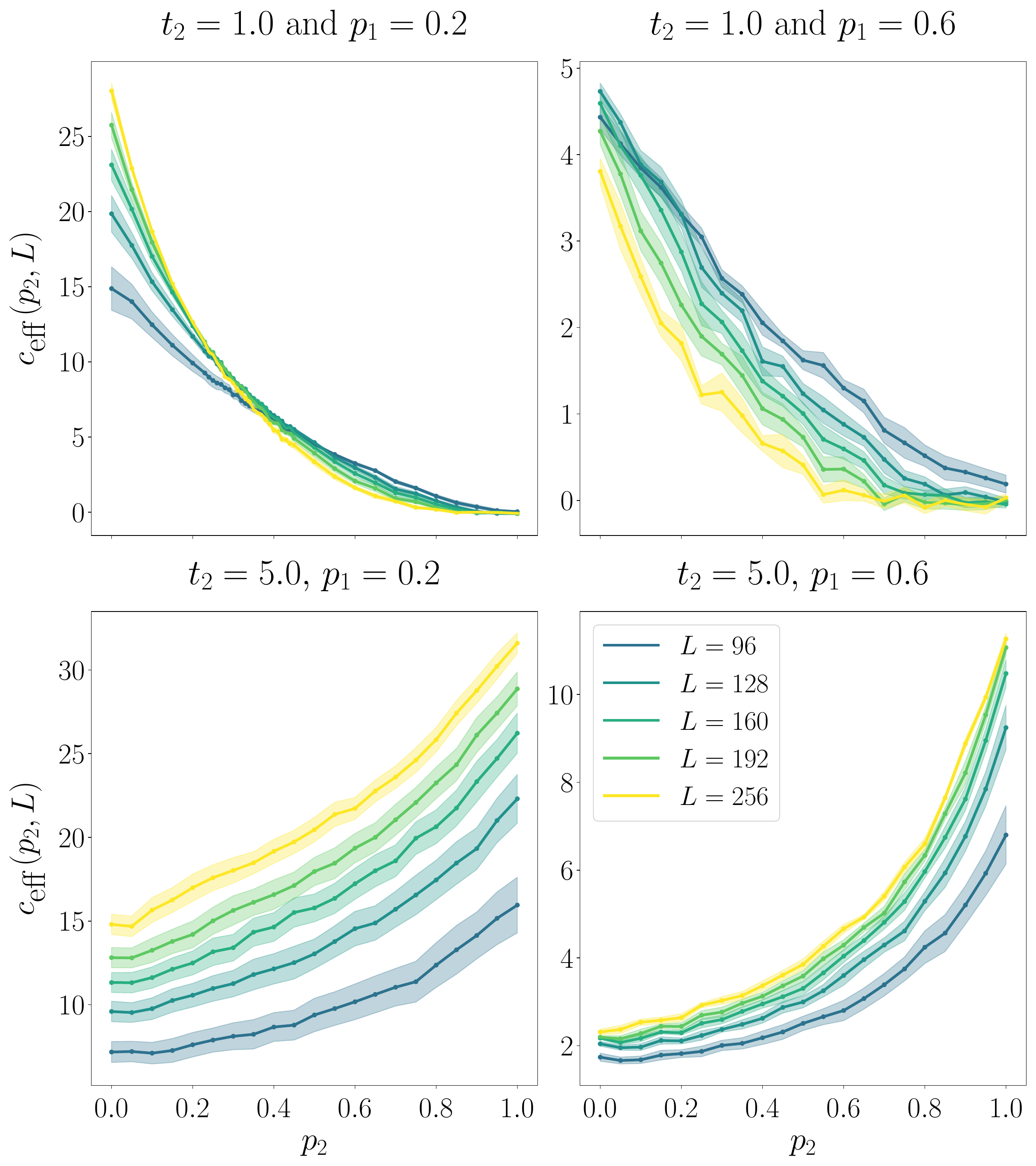}
\put(-232,268){(a)} 
\put(-112,268){(b)} 
\put(-232,137){(c)} 
\put(-112,137){(d)} 
\caption{Plot of the logarithmic prefactor $c_{\rm{eff}}$ as function of $p_2$, for various values of range of the fit indicated by the maximum size $L$ considered in the fit (each represented with a different color). Each plot represent a different value of $t_2$ and of the measurement probability in the inner chain $p_1$, similarly to Fig. \ref{fig:Neg_vs_L}. For small values of $t_2$ and $p_1$ (a), $c_{\rm{eff}}$ increases with $L$ at small $p_2$ while decreasing with $L$ at larger $p_2$, exhibiting a crossing around $p_2\sim0.25$. This behavior indicates a transition between logarithmic and area law phase as function of $p_2$. Such behavior is absent for larger $p_1$ (b), where $c_{\rm{eff}}$ always decreases with increasing $L$, or for $t_2=5$ (c)-(d) where $c_{\rm{eff}}$ always increases with $L$.}
    \label{fig:c_vs_p2}
\end{figure}

When the dynamics of the ancilla is fast ($t_2= 5.0$) we always observe a logarithmic phase, both in the weak ($p_1= 0.2$, panel c) and strong ($p_2 =0.6$, panel d) measurements on the system, and independently of the value of $p_2$. In other words, independently of the value of $p_2$, the system is never in area law -- although additional results obtained for larger values of $p_1$ show that the area law appears again as $p_1\rightarrow1$. Moreover, the system entanglement is larger for  higher values of $p_2$, consistently with what has been observed in Fig.~\ref{fig:NegL32}. Interestingly, we observe that, for the same system noise $p_1=0.2$, the negativity at large $t_2=5$ and $p_2=1$ is actually larger than the negativity at $t_2=1$ and with no ancilla noise $p_2=0$. This suggests that in the large $t_2$ regime, the ancilla noise does not only act as effective anti-noise, but also can increase correlations with respect to the slow ancilla dynamics regime.

For systems at equilibrium, it is well known that logarithmic violations of the area law are observed at criticality in the presence of conformal field theories \cite{Holzhey_1994, Calabrese_2004,Calabrese_2009}. The prefactor of the logarithmic scaling is universal and is related to the central charge $c$ of the conformal field theory. In particular, the fermionic negativity, computed for two adjacent intervals of equal length $\ell$ exhibits the scaling $\mathcal{E}=\frac{c}{4}\log \ell$ \cite{Calabrese_2012}. In analogy with the ground state of equilibrium systems, we fit the negativity according to the expression

\begin{equation}\label{Eq:Neg_fit}
\mathcal{E}_{\frac L2}(L)=\frac{c_{\rm{eff}}}{4}\log(L)+a_0
\end{equation}
where $c_{\rm{eff}}$ is the coefficient of the logarithmic scaling. We do not expect $c_{\rm{eff}}$ to hold any universal meaning, since the logarithmic scaling of the negativity is not related to a CFT, as we are investigating the steady state of an open quantum system, and not the properties of equilibrium ground states. 

The behavior of $c_{\rm{eff}}$ determines whether the system exhibits a phase transition between the logarithmic scaling phase and the area law phase, or just a crossover between the two phases. We perform the fit according to Eq. \eqref{Eq:Neg_fit} for various ranges of $L$, as detailed in Appendix \ref{App:Fit} -- each time considering larger and larger system sizes -- and extract $c_{\rm{eff}}(p_2,t_2,L)$, where $L$ indicates the largest size considered in the fitting procedure. We then study the behavior of $c_{\rm{eff}}$ as function of both $p_2$ and $t_2$.

\subsection{Transition induced by noise}\label{Sec:NoiseTrans}

In Fig.~\ref{fig:c_vs_p2} we plot the logarithmic prefactor $c_{\rm{eff}}(p_2, L)$ as function of $p_2$ for different values of $t_2$ and $p_1$, and for different system sizes $L$. The trend of $c_{\rm{eff}}$ as function of $L$ can help us determine the phase of the system in the thermodynamic limit.

In the regime of slow dynamics of the ancilla $t_2=1$, the coefficient $c_{\rm{eff}}$ always decreases with the ancilla noise $p_2$ for all values of $L$ -- i.e.  $\frac{dc_{\rm{eff}}(p_2,L)}{dp_2}<0$ -- and goes to zero for $p_2\rightarrow1$. This is in agreement with the disruptive effect of noise on the system that we expect in this regime. We also observe two different trends with respect to $L$ depending on the value of $p_1$. For low noise strength $p_1=0.2$ (see Fig~\ref{fig:c_vs_p2} panel a), $c_{\rm{eff}}$ increases with system size at low values of $p_2$ while it decreases with $L$ at higher values of $p_2$. The curves of $c_{\rm{eff}}(p_2)$ at different $L$ show a crossing near a critical probability $p_{2c}$, so that $\frac{dc_{\rm{eff}}(p_2,L)}{dL}>0$  for $p<p_{2c}$, and $\frac{dc_{\rm{eff}}(p_2,L)}{dL}>0$ for $p_2>p_{2c}$. This behavior indicates a phase transition from a logarithmic scaling phase ($c_{\rm{eff}}\neq0$) to an area law phase ($c_{\rm{eff}}=0$) occurring at $p_2=p_{2c}$. On the other hand for stronger system noise $p_1=0.6$ (reported in Fig. \ref{fig:c_vs_p2} panel b), the coefficient $c_{\rm{eff}}$ is always decreasing with system size independently of $p_2$, suggesting the presence of an area law for all values of $p_2$ in the thermodynamic limit.

In the regime of fast dynamics of the ancilla $t_2=5$, $c_{\rm{eff}}$ is always increasing with system size $L$ and with the ancilla noise $p_2$, for both $p_1=0.2$ (Fig.~\ref{fig:c_vs_p2} panel c) and $p_1=0.6$ (Fig~\ref{fig:c_vs_p2} panel d). From this observation two things can be argued: \emph{(i)} the system is always in a logarithmic scaling phase of the entanglement, even when $p_2\rightarrow1$; \emph{(ii)} the noise on the ancilla increases the entanglement content of the ancilla, effectively acting as an anti-noise that cancels the effects of the system noise $p_1$.

\begin{figure*}[t!]
    \centering
    \includegraphics[width=\textwidth]{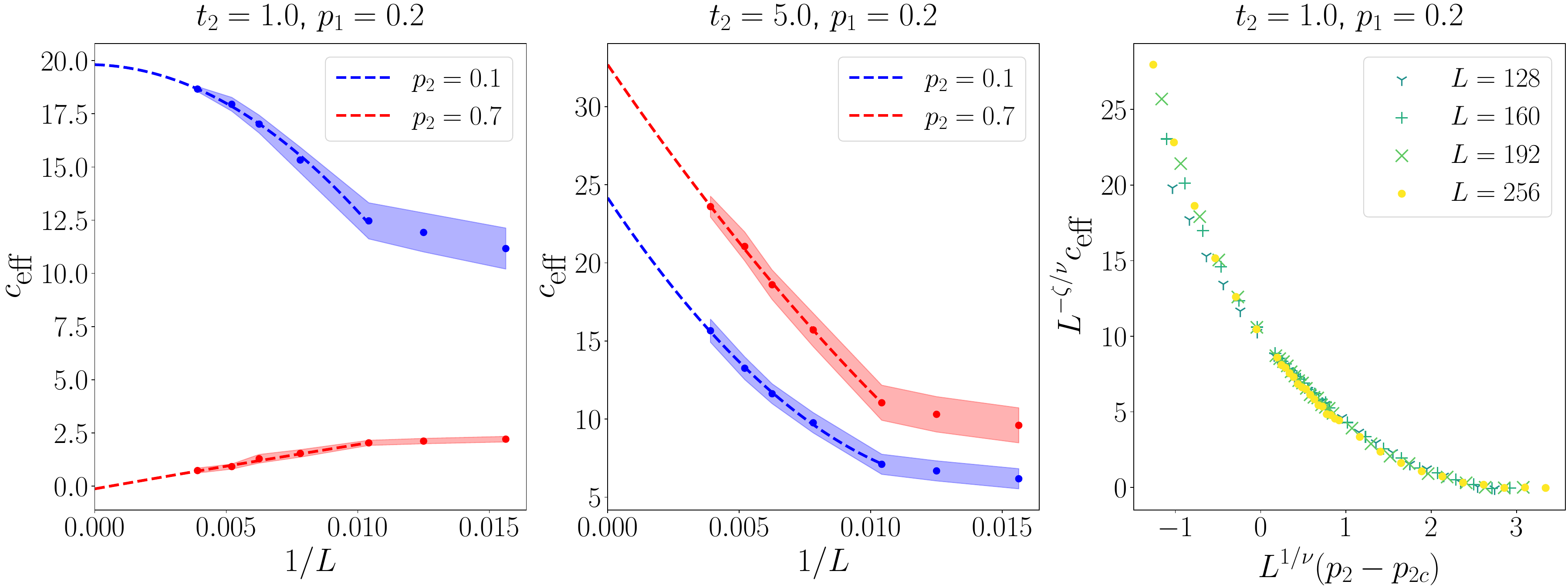}
\put(-500,184){(a)} 
\put(-330,184){(b)} 
\put(-158,184){(c)} 
    \caption{(a)-(b) Plot of $c_{\rm{eff}}$ as function of $1/L$ for $t_2=1$ (a) and $t_2=5$ (b), $p_1=0.2$ and $p_2=0.1$ (blue) $p_2=0.7$ (red). The circles represent the values of $c_{\rm{eff}}$ as obtained from the fitting procedure and the shaded region is the associated error. The dashed lines represent a polynomial fit of the second order $c_{\rm{eff}}=c_0+c_1/L+c_2/L^2$. The red curve in (a) corresponds to the area law phase, correctly showing that $c_{\rm{eff}}$ goes to zero in the thermodynamic limit $1/L\rightarrow\infty$. The other curves correspond to the logarithmic phase and exhibit a non zero value of $c_{\rm{eff}}$ in the thermodynamic limit. (c) Finite size scaling analysis of $c_{\rm{eff}}(p_2,L)$ for $t_2=1$ and $p_1=0.2$. The data show a good collapse with exponents $\nu=3.5\pm1.8$ and $\zeta=0.001\pm0.006$, and the critical probability is $p_{2c}=0.26\pm0.09$.}
    \label{fig:c_vs_1L}
\end{figure*}

The study of $c_{\rm{eff}}$ confirms the analysis from the behavior of the negativity in Fig. \ref{fig:Neg_vs_L}. These conclusions are further supported by investigating the behavior of $c_{\rm{eff}}$ in the thermodynamic limit. In Fig.~\ref{fig:c_vs_1L} we study such behavior by fitting $c_{\rm{eff}}$ as function of $1/L$ with a second order polynomial. For fixed $p_1=0.2$, $c_{\rm{eff}}$ as function of $1/L$ shows two very distinct behaviors depending on the value of $t_2$ and $p_2$. In the slow ancilla dynamics regime $t_2=1.0$ (Fig.~\ref{fig:c_vs_1L} panel a) the value of $c_{\rm{eff}}$ extrapolated at $L=\infty$ is zero for high measurement rates of the ancilla ($p_2=0.7$); this indicates that the system is in an area law. On the other hand $c_{\rm{eff}}(L\rightarrow\infty)$ attains a nonzero value for low measurement rates ($p_2=0.1$), showing that the logarithmic scaling phase survives in the thermodynamic limit too. In the fast ancilla dynamics $t_2=5.0$ regime (Fig.~\ref{fig:c_vs_1L} panel b), instead, $c_{\rm{eff}}$ always reaches a nonzero value, regardless of $p_2$, indicating that the system entanglement always scales logarithmically. This clearly shows how the dynamics of the ancilla chain changes the effect of measurements on the entanglement shared between partitions of the system chain.

We observe that $c_{\rm{eff}}$ represents a good order parameter for the transition between logarithmic phase and area law phase (where it vanishes). Therefore, it is interesting to perform a finite size scaling analysis \cite{FisherFSSA1972,binderMonteCarloSimulation2010} to check if the curves of $c_{\rm{eff}}(p_2,L)$ exhibit a collapse; such analysis is also useful to quantitatively determine the critical probability $p_{2c}$. We fit the behavior of $c_{\rm{eff}}(p_2,L)$ with the function $c_{\rm{eff}}(p_2,L)=L^{\zeta/\nu}f(L^{1/\nu}(p_2-p_{2c}))$, where $p_{2c}$ is the transition probability, while $\nu$ and $\zeta$ are the critical exponents. The results are reported in Fig. \ref{fig:c_vs_1L}, panel (c), where the various curves associated to different $L$ nicely collapse onto one curve, confirming once again that $p_2$ tunes a transition between logarithmic and area law. We find that $p_{2c}=0.26\pm0.09$, $\nu\approx3.5\pm1.8$ and $\zeta=0.001\pm0.006$. Except for the critical probability, both $\nu$ and $\zeta$ have large uncertainties.

We do not expect the values of the critical exponents or of the critical probability to have any universal meaning. However, we expect $\zeta=0$ which corresponds to a situation where $c_{\rm{eff}}$ does not scale with $L$ in the thermodynamic limit, i.e. $c_{\rm{eff}}(L\rightarrow\infty)$ can have finite non-zero values, a necessary condition for the existence of the logarithmic scaling phase. The obtained value of $\zeta$ is close enough to zero to be compatible with such behavior within the fit uncertainty.

Moreover we expect the value of $p_{2c}$ to depend on both $p_1$ and $t_2$. In particular $p_{2c}$ increases as $t_2$ is increased, since a larger ancilla hopping enhances the monogamy effect and strengthens the system entanglement; it also decreases with $p_1$, since a stronger Markovian noise on the system collapses the entanglement and induces the area law at lower values of $p_2$. Since $p_2$ ranges between 0 and 1, we can distinguish three different regimes depending on the value of $t_2$ and $p_1$. \emph{(i)} An \textit{Area-Log} regime where the system exhibits either logarithmic or area law phases, depending on the value of $p_2$. In such regime, the logarithmic phase exists for $p_2<p_{2c}(t_2,p_1)$ and the area law appears above this critical probability. \emph{(ii)} An \textit{Area-only} regime where the system is always in an area law for any value of $p_2$, i.e. even at $p_2=0$ the entanglement is constant in $L$. This regime occurs for small $t_2$ or for $p_1$ close enough to $1$. \emph{(iii)} A \textit{Log-only} regime where the system is always in a logarithmic phase and increasing $p_2$ increases the entanglement (which is maximum for $p_2=1$). This regime appears for large $t_2$ and for $p_1$ not too close to $1$.

\begin{figure}[h!]
    \centering
    \includegraphics[width=0.96\columnwidth]{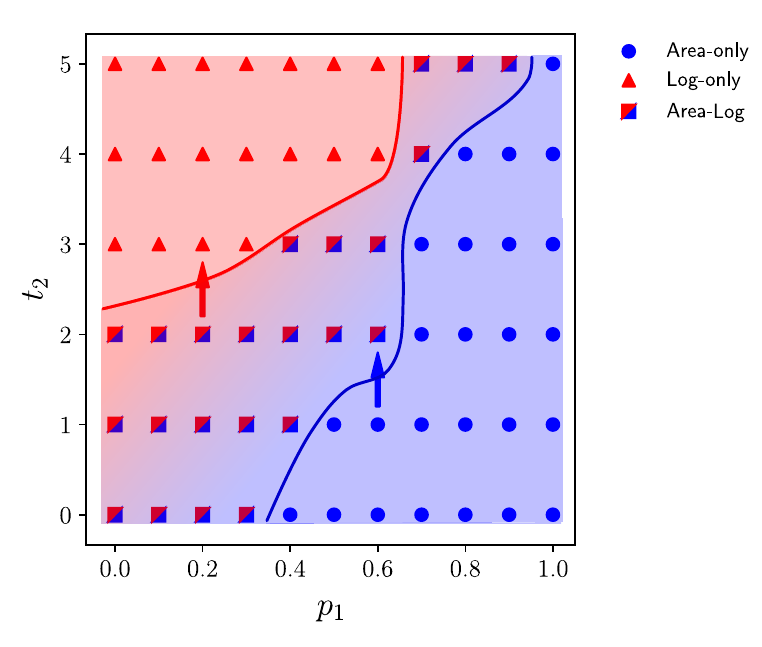}
    \caption{A sketch of the phase diagram of the system as function of the system noise strength $p_1$ and of the ancilla hopping $t_2$. The blue circles indicate that the system is always in an area law independently of the value of $p_2$. The red triangles indicate that the system entanglement always follow a logarithmic scaling even at $p_2\rightarrow1$. The red and blue squares indicate that the system undergoes a transition from logarithmic phase to area law phase as $p_2$ is increased from $p_2=0$ to $p_2=1$.}
    \label{fig:Ph_Diag}
\end{figure}

A more detailed representation of these regimes is reported in Fig. \ref{fig:Ph_Diag} where the different regimes are indicated as function of $t_2$ and $p_1$. The blue region (with blues circles) indicate the Area-only regime, and the red region (with red triangles) the Log-only regime. The red-blue shaded region (with red-blue squares) corresponds to the Area-Log regime. Of course we remark that the line $p_1=1$ is always in an area phase, since always measuring the system chain makes its entanglement equal to zero. The remarkable effect that emerges from Fig. \ref{fig:Ph_Diag} is that for regimes of fast ancilla dynamics ($t_2\gtrsim3$) the area law phase is suppressed and its appearance is pushed to large values of $p_1$. For example, while at $t_2=1$ the system enters an area law already for low noises strength ($p_1=0.2$ and $p_2=0.3$), for $t_2=5$ the area law only emerges when $p_1\gtrsim0.6$, making the logarithmic entangled phase very robust against noise on the system.

\subsection{Transition induced by hopping}\label{Sec:HoppingTrans}

In this section, we focus instead on the dependence of $c_{\rm{eff}}$ on $t_2$ at fixed $p_1$ and $p_2$. We show that $t_2$ tunes a transition between the logarithmic and area law phases, so that the hopping in the ancilla is effectively a control parameter of the transition. It is interesting to study the special cases $p_2=0$ and $p_2=1$. The first case corresponds to a transition from the Area-only regime to the Area-Log regime (blue arrow in Fig. \ref{fig:Ph_Diag}), where increasing $t_2$ induces the presence of the logarithmic phase. On the other hand, the second case corresponds to the transition from the Area-Log regime to the Log-only regime  (red arrow in Fig. \ref{fig:Ph_Diag}), where increasing $t_2$ suppresses the area phase.

\begin{figure}[t!]
    \centering
    \includegraphics[width=\columnwidth]{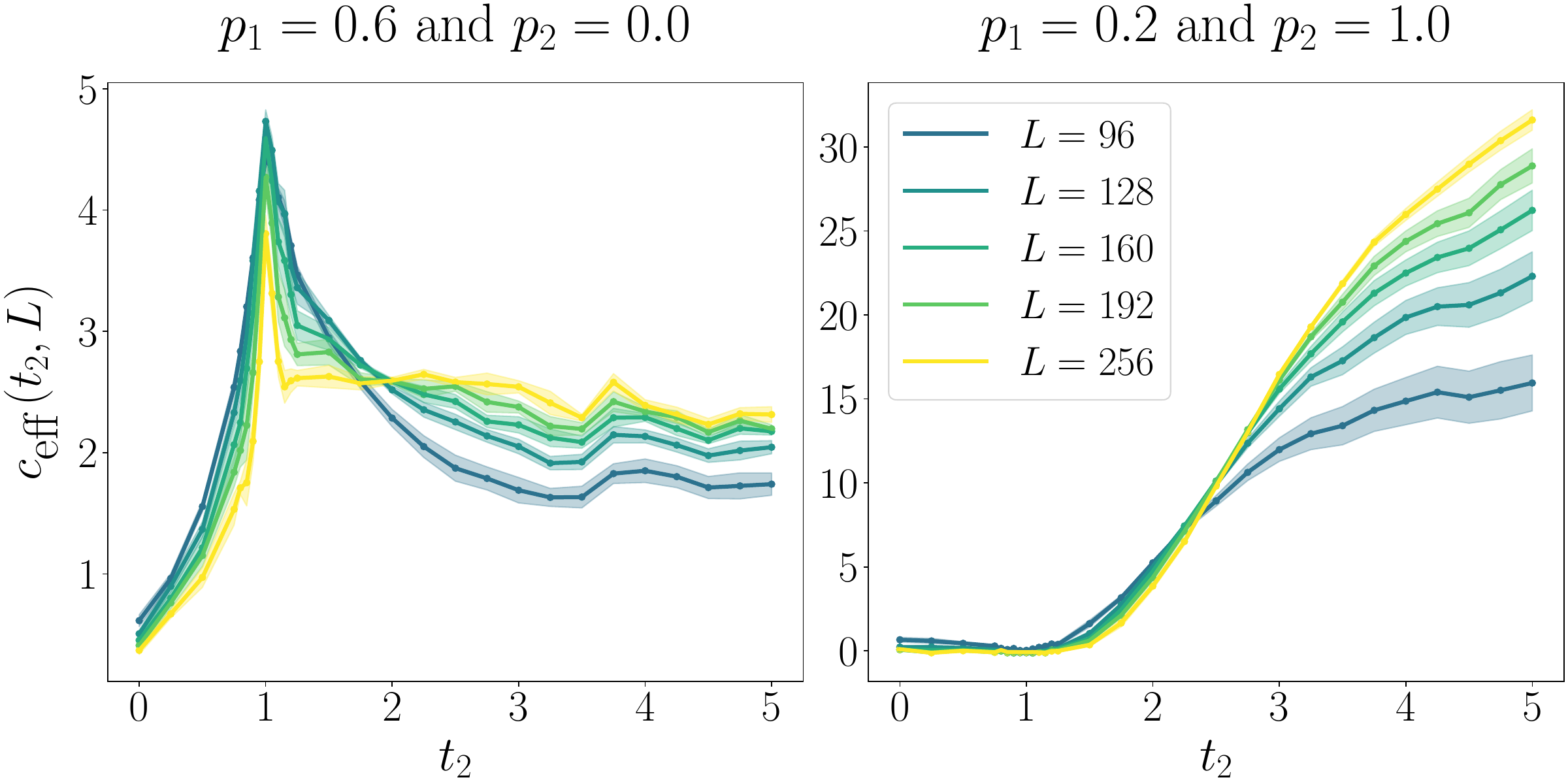}
\put(-232,115){(a)} 
\put(-112,115){(b)} 
\caption{Plot of the logarithmic prefactor charge $c_{\rm{eff}}$ as function of $t_2$, for various values of range of the fit indicated by the maximum size $L$ considered in the fit (each represented with a different color). Panel (a) reports the results for $p_1=0.6$ and $p_2=0$, which corresponds to the transition from the Area-only regime to the Area-Log regime, marked with the appearance of a logarithmic phase, indicated by the crossing near $t_2=2$. Panel (b) reports the results for $p_1=0.2$ and $p_2=1$, which corresponds to the transition from the Area-Log regime to the Log-only regime, marked with the disappearance of the area phase, indicated by the crossing near $t_2=2.5$.}
    \label{fig:c_vs_t2}
\end{figure}

We focus on $p_1=0.2$ and $p_1=0.6$ and in Fig. \ref{fig:c_vs_t2} report the behavior of $c_{\rm{eff}}(L)$ as function of $t_2$. For $p_1=0.6$ and $p_2=0$ (panel (a) of Fig. \ref{fig:c_vs_t2}), $c_{\rm{eff}}$ decreases with $L$ for $t_2<t_{2c}$ and increases with $L$ for $t_2>t_{2c}$, where $t_{2c}\approx2$ is the critical hopping where the curves exhibit a crossing. Similarly for  $p_1=0.2$ and $p_2=1$ (panel (b) of Fig. \ref{fig:c_vs_t2}), $c_{\rm{eff}}$ decreases with $L$ for $t_2<t_{2c}$ and increases with $L$ for $t_2>t_{2c}$, where $t_{2c}\approx2.5$ in this case. 

The presence of a crossing point between the curves at different values of $L$ confirms that $t_2$ induces a transition between area and logarithmic phases, with the latter always appearing at larger hopping values. 

\section{Conclusions}\label{Sec:Conclusions}

In this work we have explored how the interplay between unitary dynamics and non-Markovian noise affects the steady state value of entanglement in a free fermion system. We studied two coupled fermionic chains (a system and an ancilla), which evolve unitarily and are subject to random noise acting as projective measurements of the particle occupation number. We quantified the entanglement within the system via the fermionic negativity of two partitions of the system. We investigated how the entanglement is affected by measurements acting on the system and on the ancilla, and how it depends on the hopping parameter of the ancilla chain.

We observed two distinct regimes determined by the value of the hopping in the ancilla chain, i.e. by how fast the unitary dynamics of the ancilla is compared to the dynamics of the system. For slow dynamics of the ancilla chain, the negativity computed between partitions of the system decreases upon increasing the noise on both the system and the ancilla, and shows a transition from logarithmic scaling to an area law scaling. For fast dynamics of the ancilla, instead, the entanglement in the system decreases for stronger noise on the system, but \emph{increases} when the noise on the ancilla is stronger. 

This enhancement effect can be explained as a consequence of the monogamy of entanglement and of the memory effects associated to the ancilla noise. The monogamy principle \cite{Dur:2000monogamy,Yang2006:monogamy,Giorgi:2011monogamy} states that only a finite amount of entanglement can be shared within a system; in the fast dynamics regime, entanglement is mainly created within the ancilla and the entanglement content in the system is suppressed. A stronger ancilla noise destroys the entanglement in the ancilla, allowing it to be created within the system and thus enhancing the system entanglement. 

Another factor contributing to entanglement enhancement is due to the memory effects introduced by the dynamics of the ancilla. When the ancilla is traced out and the reduced dynamics of the system is studied, the Markovian ancilla noise is filtered by the ancilla dynamics, leading the system to experience an effective non-Markovian noise that is correlated on timescales determined by the hopping parameter $t_2$. Non-Markovianity and memory effects have been linked to a revival of coherences, and it has been showed that long-range noise (i.e. noise that is correlated in space) can increase entanglement \cite{seetharam2021correlation,Piccitto_SciPostPhysCore.6.4.078_2023,Russomanno_PhysRevB.108.104313_2023,LumiaPhysRevResearch.6.023176_2024}. It is thus reasonable that temporal correlations of the noise can have similar effects and enhance the entanglement.

In the fast dynamics regime the entangled phase with logarithmic scaling of the negativity survives even for strong system noise. In other words, the ancilla produces a sort of anti-noise effect, which protects the system from its own decoherence. This effect is qualitatively similar to some results of Ref. \cite{niroulaErrorMitigationThresholds2024}, where the anti-noise effect protects correlations in the system; such anti-noise is however introduced mathematically at the level of the master equation, but has no direct physical implementation.

Moreover, the numerical scaling analysis of the negativity suggests that the system switches between the logarithmic scaling phase and the area law phase through a transition tuned by either the ancilla hopping or the ancilla noise strength. In Ref. \cite{Mirlin2023} it was showed that a dissipative free fermion chain does not exhibit a transition from the logarithmic phase to area law phase, but rather a very sharp crossover which can be mistaken as a transition in a numerical analysis. Whether the same phenomenology applies to our model, which has a different underlying unitary dynamics, is an open question whose answer would require extensive and quite demanding numerical resources.

Our results indicate clearly that the ancilla hopping  $t_2$ is an efficient control parameter that reverses the destructive effect of the ancilla noise, turning it from noise to an effective anti-noise experienced by the system. This effect is of great interest for error mitigation strategies, since it would allow to protect the entangled phase of the system from errors by using an ancilla that has the same size as the system. This is a clear improvement compared to error correction schemes that typically require 3--5 physical qubits for each logical qubit. It also does not require a precise knowledge of the system noise, which is instead necessary for typical error mitigation strategies \cite{caiQuantumErrorMitigation2023,niroulaErrorMitigationThresholds2024}. Finally it is easily implementable on existing physical platforms \cite{Molmer1999,bulutaQuantumSimulators2009,blattQuantumSimulationsTrapped2012,blochQuantumSimulationsUltracold2012a,georgescuQuantumSimulation2014,BRAVYI2002210,debnath2016demonstration,bruzewiczTrappedionQuantumComputing2019,Pino2021,Noel21,Pagano2020quantum,gonzalez-cuadraFermionicQuantumProcessing2023} -- in contrast to the anti-noise used in \cite{niroulaErrorMitigationThresholds2024} -- since it requires short-range hopping which can be realized with standard 2--qubits gates on a ladder geometry.

Several questions remain open. First of all, using an ancilla with a fast dynamics does not allow the same degree of control that error correction schemes have. In particular, our results show that the effect of the ancilla is to protect the entanglement from errors operating on the system, but it is not clear whether this occurs at the level of the full state of the system or if simply the ancilla keeps the system within a high-entanglement region of the Hilbert space.

Secondly, the role of interactions in both the system and ancilla remains to be understood within this framework. Interactions induce a volume law scaling of the entanglement, and it would beinteresting to understand whether the regime of fast ancilla dynamics still induce anti-noise effects and whether interactions in the ancilla are needed to achieve such regime.

Finally, our work suggests that the anti-noise effect emerges when the internal dynamics of the ancilla is fast enough. Thus it is of great importance to understand whether the ancilla chain can be replaced by an ancilla with a smaller size but with enough degrees of freedom to have a structured internal dynamics, since scaling down the size of the ancilla would have obvious advantages in terms of physical implementations and error control.

\acknowledgments{We thank Z. Bacciconi, L. Capizzi, M. Dalmonte, G. Falci, M. Fossati, G. Piccitto, D.Poletti, X. Turkeshi for insightful discussions.

M. T. thanks the Simons Foundation for supporting his Ph.D. studies through Award 284558FY19 to the ICTP. G.C. is supported by ICSC – Centro Nazionale di Ricerca in High-Performance Computing, Big Data and Quantum Computing under project E63C22001000006. G.C. and M.T. acknowledge the CINECA award under the ISCRA initiative, for the availability of high performance computing resources and support.}

\bibliography{Paper}

\clearpage

\appendix

\section{Computation of the logarithmic negativity for free fermionic systems}\label{App:Negativity}

In this section, we explain how the fermionic logarithmic negativity 
\begin{equation}
\label{eq:Fermionic_negativity_APP}
    \mathcal{E}= \ln \Tr |\rho^{R_A}_1|= \ln \Tr \sqrt{\rho^{R_A}_1(\rho^{R_A}_1)^\dagger}
\end{equation}
is computed. 

Here $\rho^{R_A}_1$ is the partial time reversal of the reduced density matrix of the system $\rho_1$  with respect to the subsystem A. Its action is better specified looking to the Majorana representation of the density matrix. We recall that the density matrix can be written as a polynomial in Majorana fermion operators ($\gamma_{2j-1}:=c_j+c^\dagger_j$, $\gamma_{2j}:= i(c_j-c^\dagger_j)$) as
\begin{equation}
    \rho_1 = \sum_{\kappa, \mu} w_{\kappa,\mu} \gamma^{\kappa_{m_1}}_{m_1}...\gamma^{\kappa_{2m_{\ell_A}}}_{2m_{\ell_A}} \gamma^{\kappa_{n_1}}_{n_1}...\gamma^{\kappa_{2n_{\ell_B}}}_{2n_{\ell_B}},
\end{equation}
where $\gamma^{0}_x= \mathbb{1}$ and $\gamma^{1}_x=\gamma_x$ are the Majorana operators corresponding to mode $x$. The summation runs on all bitstrings $\kappa=(\kappa_{m_1},...,\kappa_{2m_{\ell_A}}) \in \{0,1\}^{2\ell_A}$ $\mu = (\mu_{n_1},...,\mu_{2n_{\ell_B}}) \in\{0,1\}^{2\ell_B}$ such that $|\kappa| +|\mu|$ is even, with $|\kappa|=\sum_{j}\kappa_j$($|\mu|=\sum_{j}\mu_j$). The partial time reversal operation acts as follows ~\cite{Shapourian2017}
\begin{equation}
    \rho^{R_A}_1 = \sum_{\kappa, \mu} w_{\kappa,\mu} i^{|\kappa|}\gamma^{\kappa_{m_1}}_{m_1}...\gamma^{\kappa_{2m_{\ell_A}}}_{2m_{\ell_A}} \gamma^{\kappa_{n_1}}_{n_1}...\gamma^{\kappa_{2n_{\ell_B}}}_{2n_{\ell_B}}.
\end{equation}

Due to the quadratic form of the Hamiltonian Eq.~\ref{Eq:modelH} and the choice of particle densities as the measured operators the gaussianity of the state $\rho_1$ is preserved at any time. Importantly, the partial time reversal transpose preserves the gaussianity of the state, and since gaussian operators are closed under product also $\rho^{R_A}_1(\rho^{R_A}_1)^\dagger$ is Gaussian. The product $\rho^{R_A}_1(\rho^{R_A}_1)^\dagger$ is not a density matrix, as it is not normalized. It is thus convenient to introduce its normalized version
\begin{equation}
    \rho_\times = \frac{1}{Z_\times} \rho^{R_A}_1(\rho^{R_A}_1)^\dagger,
\end{equation}
in terms of which the expression of the fermionic negativity takes the form 
    
\begin{align}
    \mathcal{E} &=  \ln \Tr \sqrt{\rho^{R_A}_1(\rho^{R_A}_1)^\dagger}\\
    &= \ln \Tr \sqrt{\rho_\times Z_\times}\\
    &= \frac{1}{2} \ln Z_\times + \ln \Tr \sqrt{\rho_\times}
\end{align}
It is easy to show that $Z_\times = \Tr (\rho^{2}_{1})$ which makes the first term directly computable from the spectrum of the correlation matrix associated with the original density matrix $\rho_1$. The second term, instead can be computed from the spectrum of the correlation matrix associated with $\rho_\times$.

We now detail how to calculate the fermionic negativity in terms of the two-point correlation function matrix of the ladder, which is  denoted by 
\begin{equation}
\mathcal{D}_{ij,\sigma \sigma'}(\tau)= \bra{\Psi(\tau)}\hat{c}^\dagger_{i,\sigma}\hat{c}_{j,\sigma'} \ket{\Psi(\tau)}.
\end{equation}
Let's call $\mathcal{D}^{(1)}$ the correlation matrix restricted to the system. We define 
\begin{equation}
    (\Gamma^{(1)})_{ij}= 2(\mathcal{D}_1)_{ij}-\delta_{ij}.
\end{equation}
Given a bipartition of the system into subsystems $A$ and $B$ the matrix $\Gamma^{(1)}$ takes the block form
\begin{equation}
    \Gamma^{(1)} = \begin{pmatrix}
        \Gamma^{(1)}_{AA} & \Gamma^{(1)}_{AB}\\
        \Gamma^{(1)}_{BA} & \Gamma^{(1)}_{BB}
    \end{pmatrix}
\end{equation}

It is convenient to introduce the matrices 
\begin{equation}
    \Gamma_{\pm} = \begin{pmatrix}
        \Gamma^{(1)}_{AA} & \pm i \Gamma^{(1)}_{AB}\\
        \pm i\Gamma^{(1)}_{BA} & -\Gamma^{(1)}_{BB}
    \end{pmatrix}
\end{equation}

which are the correlation matrices associated with $\rho^{R_A}_1$ and $(\rho^{R_A}_1)^\dagger$.
The fermionic negativity can be computed out of the spectrum $\{\lambda_j\}$ of $\mathcal{D}_1$ and the spectrum $\{\mu_j\}$ of the matrix $\Gamma^{(1)}_\times$\footnote{The matrix $\Gamma^{(1)}_\times$ does not actually denote the correlation matrix associated with $\rho_\times$, but is isospectral to it \cite{EisertNegativity}} defined as \cite{Fagotti_2010,EisertNegativity}
\begin{equation}
    \Gamma^{(1)}_\times=\frac{1}{2}[1-(1+\Gamma^{(1)}_+\Gamma^{(1)}_-)^{-1}(\Gamma^{(1)}_+ + \Gamma^{(1)}_-)],
\end{equation}
in particular it holds \cite{Shapourian_2019}
\begin{equation}
    \mathcal{E}_A = \sum^{L}_{j=1}\biggr[\log(\sqrt{\mu_j}+\sqrt{1-\mu_j})+\frac{1}{2}\log((1-\lambda_\alpha)^2+\lambda^2_\alpha)\biggr]
\end{equation}

\section{Evolution of the two-point correlation matrix}\label{App:Evolution}
In this section we summarize how the interplay between unitary dynamics and measurements affect the evolution of the correlation matrix of the fermionic ladder. We briefly summarize the dynamics 
\begin{itemize}
    \item Unitary dynamics $\ket{\Psi} \mapsto U \ket{\Psi(\tau_u)}$ generated by the hamiltonian Eq. \ref{Eq:modelH} $U=e^{-i\tau_u H}$.
    \item Measurement of the fermionic number $\hat{n}_{i,\sigma}$ on each site with probability $p_\sigma$, $\sigma=1,2 $
    
\end{itemize}
All the spectra relevant to the computation of the fermionic negativity can be obtained through the spectrum of the correlation matrix reduced to the system's degrees of freedom. Therefore, to compute the negativity at any time, it is sufficient to know how the correlation matrix $\mathcal{D}(\tau)$ evolves.
During the unitary part of the evolution, the correlation matrix changes according to \cite{Coppola2022}
\begin{equation}
        \mathcal{D}(\tau+\tau_u) = \hat{\mathbb{R}}^{\dagger}\mathcal{D}(\tau)\hat{\mathbb{R}};
        \quad
        \hat{\mathbb{R}}=\frac1L\sum_ke^{-ik(m-n)}\hat{U}_{k},
\end{equation}
with $\hat{U}_{k}$ defined in Eq.(\ref{U_k}).

For what concerns the effect of measurements, since the operators $n_{l,\mu}$ and $1-n_{l,\mu}$ are othogonal projectors we have that the probability to measure $n_{l,\mu}=1$ is given by $p_{n_{l,\mu}=1}(t)=\bra{\Psi(\tau)}n_{l,\mu} \ket{\Psi(\tau)}$ and $p_{n_{l,\mu}=0}(\tau)=1-p_{n_{l,\mu}=1}(\tau)$. As a consequence the effect of density measurements translates in the following update rule for the correlation matrix $\mathcal{D}_{ij,\sigma\sigma'}(\tau)$:
\begin{enumerate}
    \item for each site $i$ belonging to chain $\mu$ extract a random number $z_{l,\mu} \in (0,1] $. If $z_{l,\mu} \leq p_\mu$ the
measurement is performed. 
\item if the measurement has to be performed extract a second random number $q_{l,\mu} \in (0,1]$.
\item If $q_{l,\mu} \leq \langle n_{l,\mu} \rangle(\tau)$  the operator $\hat{n}_{l,\mu}$ is applied to the state $\ket{\Psi(\tau)} \mapsto \hat{n}_{l,\mu}\ket{\Psi(\tau)}/||\hat{n}_{l,\mu}\ket{\Psi(\tau)}||$ which, applying to Wick theorem, results into a change of $\mathcal{D}_{ij,\sigma\sigma'}(\tau)$ given by 
\begin{equation}
    \begin{split}
        \mathcal{D}_{ij,\sigma \sigma'}(\tau) & \to \mathcal{D}_{ij,\sigma \sigma'}(\tau) + \delta_{il}\delta_{jl}\delta_{\sigma \mu}\delta_{\sigma' \mu} 
        \\
        &- \frac{\mathcal{D}_{il,\sigma \mu}(\tau)\mathcal{D}_{lj,\mu \sigma'}(\tau)}{\mathcal{D}_{ll,\mu\mu}(\tau)}.
    \end{split}
\end{equation}
\item If $q_{l,\mu} > \langle n_{l,\mu} \rangle(\tau)$, then the operator $1 - \hat{n}_{l,\mu}$ is applied to the state, $\ket{\Psi} \mapsto (1-\hat{n}_{l,\mu})\ket{\Psi}/||(1-\hat{n}_{l,\mu})\ket{\Psi}||$ which results into into a change of $\mathcal{D}_{ij,\sigma\sigma'}(\tau)$ given by 
\begin{equation}
    \begin{split}
        &\mathcal{D}_{ij,\sigma \sigma'}(\tau)  \to
        \mathcal{D}_{ij,\sigma \sigma'}(\tau) - \delta_{il}\delta_{jl}\delta_{\sigma \mu}\delta_{\sigma' \mu} 
        \\
        & + \frac{(\delta_{il,\sigma \mu}-\mathcal{D}_{il,\sigma \mu}(\tau))(\delta_{lj,\mu\sigma'}-\mathcal{D}_{lj,\mu \sigma'}(\tau))}{(1-\mathcal{D}_{ll,\mu\mu}(\tau))}.
    \end{split}
\end{equation}

\end{enumerate}

\section{Details of the fitting procedure}\label{App:Fit}

We perform a linear regression fit for $\overline{\mathcal{E}}_{\frac{L}{2}}$ as function of $\log L$. In order to capture how the coefficient of the logarithm drifts in the thermodynamics limit, we employ moving fitting intervals in $L$. 

The five fitting ranges we consider are

\begin{itemize}
    \item $L\in(8,16,24,32,48,64,80,96)$
    \item $L\in(16,24,32,48,64,80,96,128)$
    \item $L\in(24,32,48,64,80,96,128,160)$
    \item $L\in(32,48,64,80,96,128,160,192)$
    \item $L\in(48,64,80,96,128,160,192,256)$
\end{itemize}

each identified by the maximum value of $L$ in the range. This protocol reduces the accuracy of the fit at lower values of $L$, but increases it at larger values of $L$, for which the fit is performed in a range where finite size effects are smaller. 

The uncertainty on the fitting parameters is obtained with standard techniques.

\end{document}